\documentclass[aps,prd,twocolumn,showpacs,floatfix,preprintnumbers,amsfont,amsmath,amssymb,nofootinbib, superscriptaddress]{revtex4}

\usepackage{color}
\usepackage[font=small]{caption,subfig}
\usepackage{hyperref}
\usepackage[normalem]{ulem}
\usepackage{lipsum}
\usepackage{url}
\usepackage{float}
\usepackage{ctable}
\usepackage{graphicx}
\usepackage[font=small,
   justification=justified,
   format=plain]{caption} % 'format=plain' avoids hanging indentation

\newcommand{\liang}[1]{ {\color{magenta} #1} }

\newcommand{\sk}[1]{}

\newcommand{\refeq}[1]{Eq.~(\ref{eq:#1})}          
\newcommand{\refeqs}[2]{Eqs.~(\ref{eq:#1})--(\ref{eq:#2})}

\def\VEV#1{\langle #1 \rangle}

\def\bin{{\rm b}}

\newcommand{\be}{\begin{equation}}
\newcommand{\ee}{\end{equation}}
\newcommand{\ba}{\begin{eqnarray}}
\newcommand{\ea}{\end{eqnarray}}

\newcommand{\pasp}{Publ. Astron. Soc. Pac.}

\allowdisplaybreaks
\begin{document}

\title{Relative Binning and Fast Likelihood Evaluation for Gravitational Wave Parameter Estimation}

\author{Barak Zackay}
\email{bzackay@ias.edu}
\affiliation{\mbox{School of Natural Sciences, Institute for Advanced Study, 1 Einstein Drive, Princeton, New Jersey 08540, USA}}
\author{Liang Dai}
\thanks{NASA Einstein Fellow}
\affiliation{\mbox{School of Natural Sciences, Institute for Advanced Study, 1 Einstein Drive, Princeton, New Jersey 08540, USA}}
\author{Tejaswi Venumadhav} 
\affiliation{\mbox{School of Natural Sciences, Institute for Advanced Study, 1 Einstein Drive, Princeton, New Jersey 08540, USA}}

\date{\today}

%%%%%%%%%%%%%%%%%%%%%%%%%%%%%%%%%%%%%%%%%%%%%%%%%%%%%%%%%%%%%%%%%%%%%%%%%%%%%%%

\begin{abstract}

We present a method to accelerate the evaluation of the likelihood in gravitational wave parameter estimation. Parameter estimation codes compute likelihoods of similar waveforms, whose phases and amplitudes differ smoothly with frequency. We exploit this by precomputing frequency--binned overlaps of the best--fit waveform with the data. We show how these summary data can be used to approximate the likelihood of any waveform that is sufficiently probable within the required accuracy. We demonstrate that $\simeq 60$ bins suffice to accurately compute likelihoods for strain data at a sampling rate of $4096\,$Hz and duration of $T=2048\,$s around the binary neutron star merger GW170817\sk{ \liang{[Liang: Shall we instead/also say ``unbiased parameter estimation''?]}}. Relative binning speeds up parameter estimation for frequency domain waveform models by a factor of $\sim 10^4$ compared to naive matched filtering and $\sim 10$ compared to reduced order quadrature.

\end{abstract}

%%%%%%%%%%%%%%%%%%%%%%%%%%%%%%%%%%%%%%%%%%%%%%%%%%%%%%%%%%%%%%%%%%%%%%%%%%%%%%%

\maketitle

%%%%%%%%%%%%%%%%%%%%%%%%%%%%%%%%%%%%%%%%%%%%%%%%%%%%%%%%%%%%%%%%%%%%%%%%%%%%%%%

%%%%%%%%%%%%%%%%%%%%%%%%%%%%%%%%%%%%%%%%%%%%%%%%%%%%%%%%%%%%%%%%%%
%\section{Introduction}
%%%%%%%%%%%%%%%%%%%%%%%%%%%%%%%%%%%%%%%%%%%%%%%%%%%%%%%%%%%%%%%%%%

{\it Introduction: }Parameter estimation is a crucial step in extracting astrophysical information from gravitational wave (GW) signals. This process involves repeatedly evaluating the likelihood of the data with waveforms that are small perturbations from the best-fit waveform (as opposed to detection). A na\"{i}ve computation of the likelihood requires a fine grid of frequencies, since typical gravitational waveforms oscillate rapidly with frequency. This process is computationally expensive, especially when analyzing long chunks of strain data with high sampling rates. Significant efforts have been invested in improving the parameter estimation methods, primarily in order to reduce the runtime of waveform generation and matched filtering to the data ~\cite{Canizares:2014fya, Miller:2015sga, Moore:2015sza, Purrer:2015tud, OShaughnessy:2017tak}. 

In this letter, we present a conceptually and practically simple way to speed up likelihood evaluations for GW signals. The key idea is that sampled waveforms with non-negligible posterior probability are all very similar to each other in the frequency domain, and differ only by smoothly varying perturbations. If we work directly with the ratios of the candidate and fiducial waveforms in the frequency domain, we can operate with a lower resolution without losing accuracy in computing the likelihood (and hence the name `relative binning').

\begin{figure}[t]
\begin{center}
  \includegraphics[width=\columnwidth]{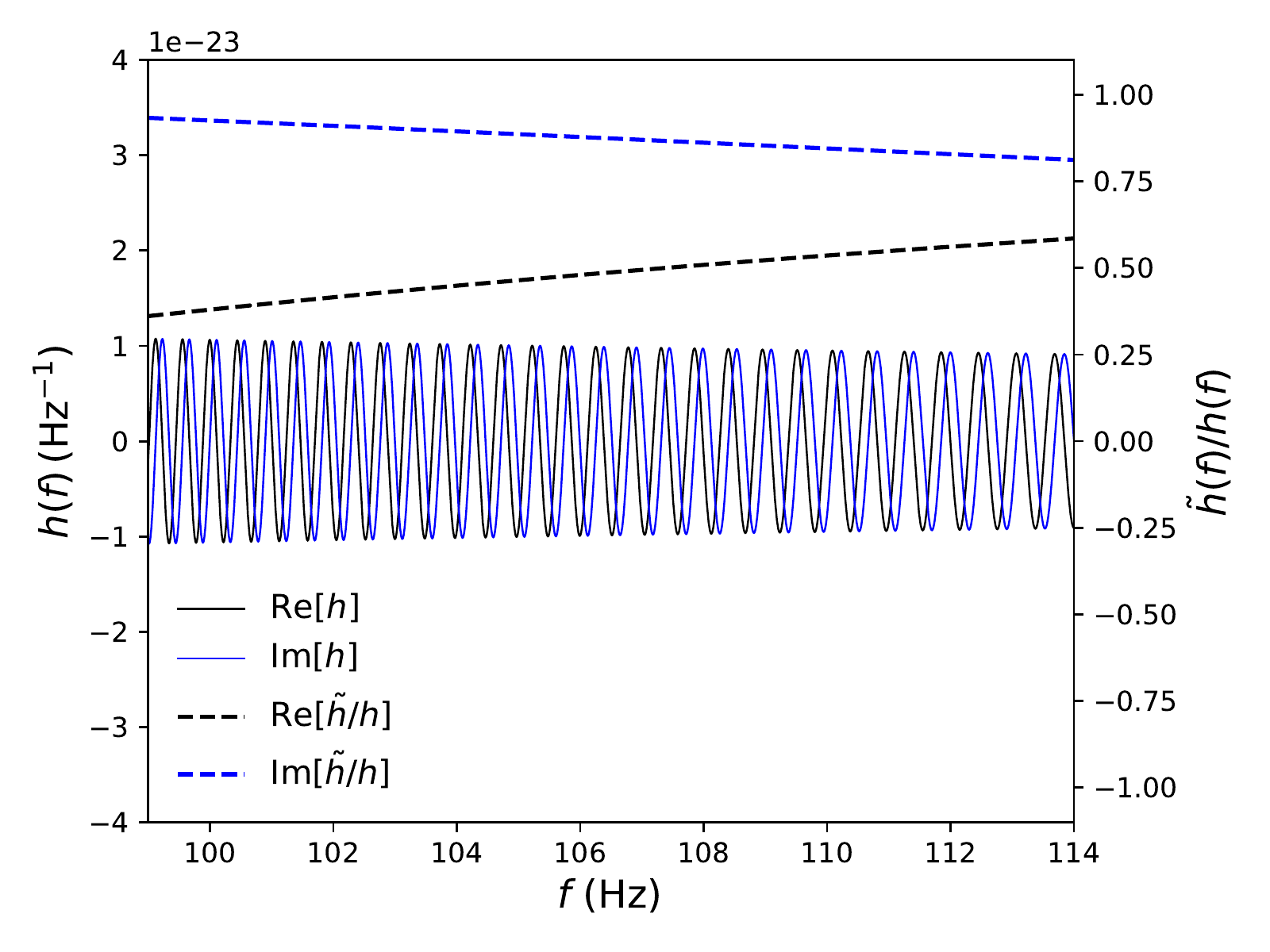}
  \caption{\label{fig:smoothdiff} Solid lines show the frequency domain waveform for an equal-mass binary neutron-star merger with chirp mass (detector frame) $\mathcal{M}^{\rm det} = 1.1975 \, M_\odot$. The dashed lines show the ratio between the perturbed ($\tilde{h}$) and the original ($h$) waveforms when the chirp mass increases by $10^{-3} \, M_\odot$, keeping all other parameters fixed. The range of frequencies corresponds to one bin in our binning scheme. We observe that this ratio varies smoothly with frequency, and is very well approximated by a linear function.}
\end{center}
\end{figure}

Figure \ref{fig:smoothdiff} shows the waveforms of two binary neutron-star mergers that differ in the detector-frame chirp mass by one part in a thousand; Even this tiny change to the parameters reduces the inner product of the two waveforms to such an extent that a converged posterior sampler would never have to sample both of them. We see that even the ratio between such widely separated waveforms varies smoothly with frequency (unlike the waveforms themselves).

%%%%%%%
%%%%%%%

We exploit this similarity by computing summary data in very coarse frequency bins for a fiducial waveform, which we choose to be the one that maximizes the likelihood, although in principle it can be any waveform that closely resembles the best-fit solution. 
We show that the summary data can be used to accurately compute the likelihood of any waveform that is sufficiently close to the fiducial one.
Finally, we demonstrate that this dramatically reduces the number of frequency bins required to compute the data likelihood for a neutron-star merger event by approximately four orders of magnitude compared to the naive computation.
Existing solutions for fast likelihood computation such as multiband interpolation and reduced order quadrature (Refs.~\cite{Vinciguerra:2017ngf, Smith:2016qas}) try to capture the structure of all possible waveforms in some broad category.
Despite their great ingenuity, those solutions still require an order of magnitude more summary statistics than relative binning does. This is primarily because waveforms with different numbers of cycles in the detector sensitive band are nearly orthogonal.
The idea of summarizing the data relative to some fiducial waveform was previously presented in Ref.~\cite{2000PhRvD..62h2001T}, in a different context (efficient detection of compact binary mergers using matched filtering). The unequal sizes of the frequency bins, the method of choosing the bins, and the linear approximation within, are all novel and crucial for accelerating parameter estimation. Together, these reduce the required number of bins, and consequently speed up parameter estimation by more than an order of magnitude.
%%%%%%%%%%%%%%%%%%%%%%%%%%%%%%%%%%%%%%%%%%%%%%%%%%%%%%%%%%%%%%%%%%
%\section{Relative binning}
%%%%%%%%%%%%%%%%%%%%%%%%%%%%%%%%%%%%%%%%%%%%%%%%%%%%%%%%%%%%%%%%%%

{\it Relative binning:} 
Denoting by $d(f)$ the signal measured in one gravitational wave detector, $h(f)$ the true waveform of the GW source, and $n(f)$ the detector noise described by a stationary Gaussian random variable with a one-sided power spectrum density (PSD) $S_n(f)$, we have $d(f) = h(f) + n(f)$. We adopt the convention of the discrete Fourier transform---the frequency $f$ takes discrete values from the reciprocal grid of a time-domain sequence of length $T$, and all frequency-domain strains are dimensionless. For simplicity, we only consider the dominant quadrupolar ($\ell=2$) components of the gravitational waves.

Assume that we only have to compute the likelihood for ``similar'' waveforms $h(f)$ such that the complex-valued ratio $r(f) = h(f)/h_0(f)$ is smooth in frequency, where $h_0(f)$ is the fiducial waveform. Inside a single frequency bin $\bin=[f_{\rm min}(\bin), f_{\rm max}(\bin)]$, a linear interpolation is a good approximation if the bin width is sufficiently small, i.e.,
\ba
\label{eq:rf}
r(f) = \frac{h(f)}{h_0(f)} = r_0(h,\bin) + r_1(h,\bin)\,(f - f_{\rm m}(\bin)) + \cdots,
\ea
where $f_{\rm m}(\bin)$ is the central frequency of the bin and terms of $\mathcal{O}\left[(f - f_{\rm m}(\bin))^2\right]$ are neglected. In practice, the bin coefficients $r_0(h, \bin)$'s and $r_1(h, \bin)$'s can be efficiently derived from the values of $r(f)$ at the bin edges. Evaluating the likelihood requires us to compute the following generic complex-valued overlaps:
\begin{align}
\label{eq:Z}
Z\left[ d(f), h(f) \right] \equiv 4 \sum_f \frac{d(f)\,h^*(f)}{S_n(f)/T}.
\end{align}
We need both the real part and the imaginary part of $Z[d(f), h(f)]$ to analytically marginalize or maximize with respect to an overall $f$-independent wave phase $\phi_c$. Similarly, we need to compute $Z[h(f), h(f)]$ to analytically marginalize or maximize with respect to an unknown normalization of the waveform.

We can efficiently compute $Z[d(f), h(f)]$ and $Z[h(f), h(f)]$ for a large number of sampled $h(f)$'s by preparing the following {\it summary data}:
\ba
\label{eq:A0}
A_0(\bin) & = & 4\sum_{f\in \bin}{\frac{d(f)\,h^*_0(f)}{S_n(f)/T}}, \\
\label{eq:A1}
A_1(\bin) & = & 4\,\sum_{f\in \bin}{\frac{d(f)\,h^*_0(f)}{S_n(f)/T}}\,(f - f_{\rm m}(\bin)), \\
B_0(\bin) & = & 4\,\sum_{f\in \bin}{\frac{|h_0(f)|^2}{S_n(f)/T}}, \\
B_1(\bin) & = & 4\,\sum_{f\in \bin}{\frac{|h_0(f)|^2}{S_n(f)/T}}\,(f - f_{\rm m}(\bin)).
\ea
These summary data are computed at maximal resolution.
For a given test waveform $h(f)$, we only need to supply one pair of coefficients $r_0(h, \bin)$ and $r_1(h, \bin)$ for each frequency bin, which encode the deviation of $h(f)$ from the fiducial $h_0(f)$. The advantage is that the total number of evaluations needed to compute the coefficients $r_0(h, \bin)$ and $r_1(h, \bin)$ is small, on the order of the number of frequency bins, when a frequency-domain waveform model is available.

We can then compute the complex-valued overlaps using
\ba
Z[d(f), h(f)] & \approx & \sum_{\bin}\,\Big( A_0(\bin)\,r^*_0(h, \bin) + A_1(\bin)\,r^*_1(h, \bin) \Big), \nonumber\\
Z[h(f), h(f)] & \approx & \sum_{\bin}\,\Big( B_0(\bin)\,\left|r_0(h, \bin)\right|^2  \nonumber\\ 
&& + 2\,B_1(\bin)\,\mathfrak{Re}[r_0(h, \bin)\,r^*_1(h, \bin)] \Big),
\ea
achieving an accuracy at linear order in $(f-f_{\rm m}(\bin))$.

{\it Choice of binning scheme:} We now discuss how to choose a minimal set of frequency bins in order to achieve a given tolerance on the accuracy of likelihood evaluation.

\begin{table}
\centering
\caption{A comparison between the computational complexity of different proposals for fast computation of the likelihood function. We consider a NS-NS merger event similar to GW170817.
\label{Table1}}
\begin{tabular}{ |c|c| }
\hline
  Method & Waveform Evaluations \\ 
 \hline
 \hline
 Relative binning [this work] & 63 \\
 \hline 
 Reduced order quadrature~\cite{Smith:2016qas} & 1740 \\
 \hline
 Multi-band interpolation~\cite{Vinciguerra:2017ngf}\footnote{We note that the authors suggest up-sampling to the full FFT grid while computing the complex overlaps.} & $\sim 10^4$ \\
 \hline
 Full FFT grid & $\sim 10^{7}$ \\
 \hline
 
\end{tabular}
\end{table}
%%%%%%%%%%%%%%%%%%%%%%%%%%%%%%%%%%%%%%%%%%%%%%%%%%%%%%%%%%%%%%%%%%%
Consider the phase $\Psi(f) = {\rm arg}[h(f)]$ expressed in terms of a sum of power laws
\ba 
\label{eq:Psif}
\Psi(f) = \sum_i\,\alpha_i\,f^{\gamma_i},
\ea
where each coefficient $\alpha_i$ encodes the effect of one or a few of the intrinsic or extrinsic parameters of the GW source. The form of \refeq{Psif} is motivated by post-Newtonian theory~\cite{2014LRR....17....2B}. For instance, the chirp mass $\mathcal{M}^{\rm det}$ enters from the leading order with $\gamma_i = -5/3$, the symmetric mass ratio $\eta$ enters from the 1PN order with $\gamma_i = -1$, and the leading order effect of spins has $\gamma_i = -2/3$. For compact stars, tidal deformation enters with $\gamma_i = 5/3$ and beyond. Moreover, a change in the merger time corresponds to phase shifts with $\gamma_i = 1$. 

In order for the likelihood to remain substantial after parameters are perturbed from the best-fit solution, a single term in \refeq{Psif} should not vary by more than a few phase cycles. This allows the absolute value of the coefficient $\alpha_i$ to change by at most $\delta\alpha^{\rm max}_i \approx 2\pi\,\chi/(f_{*,i})^{\gamma_i}$, where we introduce a tunable factor $\chi$. We choose the characteristic frequency to be $f_{*,i} = f_{\rm min}$ for $\gamma_i < 0$ and $f_{*,i} = f_{\rm max}$ for $\gamma_i > 0$, where $[f_{\rm min}, \,f_{\rm max}]$ is the frequency range with a significant contribution to the overall matched filter signal-to-noise ratio. Accounting for a number of terms in \refeq{Psif}, a maximum phase change would be
\ba 
\delta\Psi_{\rm max}(f) & = & \sum_i\, \delta\alpha_i^{\max}\, f^{\gamma_i}\,{\rm sgn}(\gamma_i) \nonumber\\
& = & 2\pi\,\chi\,\sum_i\,\left(f/f_{*,i}\right)^{\gamma_i}\,{\rm sgn}(\gamma_i),
\ea
where we have added a factor of ${\rm sgn}(\gamma_i)$ to account for the worst-case scenario in which the signs of $\alpha_i$'s conspire to produce the maximum possible {\it differential} phase change with frequency. Thus, we can divide $[f_{\rm min}, \,f_{\rm max}]$ into bins based on the criterion that the differential phase change within each bin $f \in [f_{\rm min}(\bin), \,f_{\rm max}(\bin)]$ is no more than some small number $\epsilon$ in radians,
\ba
\left| \delta\Psi_{\rm max}\left(f_{\rm max}(\bin)\right) - \delta\Psi_{\rm max}\left(f_{\rm min}(\bin)\right) \right| & < & \epsilon. 
\ea
This algorithm for determining the bins does not require detailed physics knowledge about the values of the $\alpha_i$'s.

\textit{Comparison to other works:}
A good metric for comparing our method to past works is the number of frequency domain waveform evaluations required to compute the likelihood to sufficient precision. Table~\ref{Table1} presents such a comparison.
We denote by $\mathcal{L}(h(f))$ the likelihood of the data given $h(f)$, we further denote the difference between computing the log-likelihood exactly and computing it using the binned data by $\Delta \ln \mathcal{L}_{\rm bin}(h(f))$. Assuming that we have picked the fiducial waveform $h_0(f)$ to be the maximum likelihood waveform, we denote it's likelihood by $\mathcal{L}_{\max}$.
Our method naturally produce likelihood errors that could be expressed using
\ba 
\label{eq:like-err}
\Delta \ln \mathcal{L}_{\rm bin}(h(f)) \approx \beta \, |\ln\mathcal{L}_{\max} - \ln\mathcal{L}(h(f))|
\ea
because both the log-likelihood difference and the binning error scale quadratically with the amplitude of any phase perturbation.
We aim for $\beta <0.01$, which is similar to that adopted in Ref.~\cite{Vinciguerra:2017ngf}).
This requires $62$ bins, and hence $124$ complex multiplications to apply for computing $Z[d(f),h(f)]$.
If we approximate the likelihood function locally as a multivariate Gaussian, this implies that the approximation bias to any inferred parameter is less than $\approx \beta=1\%$ of its own standard deviation. Moreover, other effects, such as the finite size of the independent samples sampled from the posterior distribution, produce comparable errors. Note that in principle $Z[h(f),h(f)]$ requires substantially fewer bins compared to $Z[d(f),h(f)]$, and is therefore disregarded from the complexity analysis.

We comment that Ref.~\cite{Smith:2016qas} aimed at much more stringent levels of accuracy in representing the waveforms (up to $5\times 10^{-12}$ in terms of the relative accuracy $\approx \Delta \ln \mathcal{L}/\ln\mathcal{L}$). By examining Fig 4. in Ref.~\cite{Smith:2016qas}, we see that reducing the accuracy goal to a more moderate value (such as the criterion we adopt here, which is sufficient for any practical purpose) does not proportionally reduce the dimension of the basis. To achieve acceptable accuracy reduced order quadrature would still need more than $10^3$ components. We can intuitively understand this as the consequence of the thousands of cycles that neutron star mergers have within the LIGO band, to accurately capture which requires roughly $\sim 10^3$ independent components.

If accuracy needs to be improved, two possible strategies can be adopted. One way is to simply increase the number of bins. Alternatively, one can extend \refeq{rf} to polynomials of higher degree, in which case the procedure to construct and apply summary data generalizes in a straightforward manner. We find that if the phase error within each bin is smaller than one radian, it is preferable to increase the order, and that if it is larger then more bins are required.  

We apply the binning technique to the parameter estimation for the double neutron-star (NS) event GW170817~\cite{abbott2017gw170817} using the publicly available strain data from both LIGO detectors~\cite{2015JPhCS.610a2021V}. This event requires the finest matched-filter search in the parameter space among all LIGO detections to date, and thus parameter estimation is computationally much more demanding than in the binary black hole case.

The relative binning technique allows us to analyze the full $T=2048\,$s stream of cleaned strain data at a sampling rate of $4096\,$Hz. We coupled our likelihood code to the Markov-chain Monte-Carlo (MCMC) ensemble sampler \texttt{emcee}~\cite{goodman2010ensemble, 2013PASP..125..306F} and generated estimates of the posterior distributions of the parameters. We adopt the same waveform model and priors as in the latest analysis presented by LIGO~\cite{Abbott:2018wiz}, and obtain statistically similar posteriors. We present the final inferred posteriors, and technical details regarding waveform generation, sampling strategy, choice of parameters, and priors in a separate note~\cite{GW170817note}.

%%%%%%%
\begin{figure}[t]
\begin{center}
  \includegraphics[width=\columnwidth]{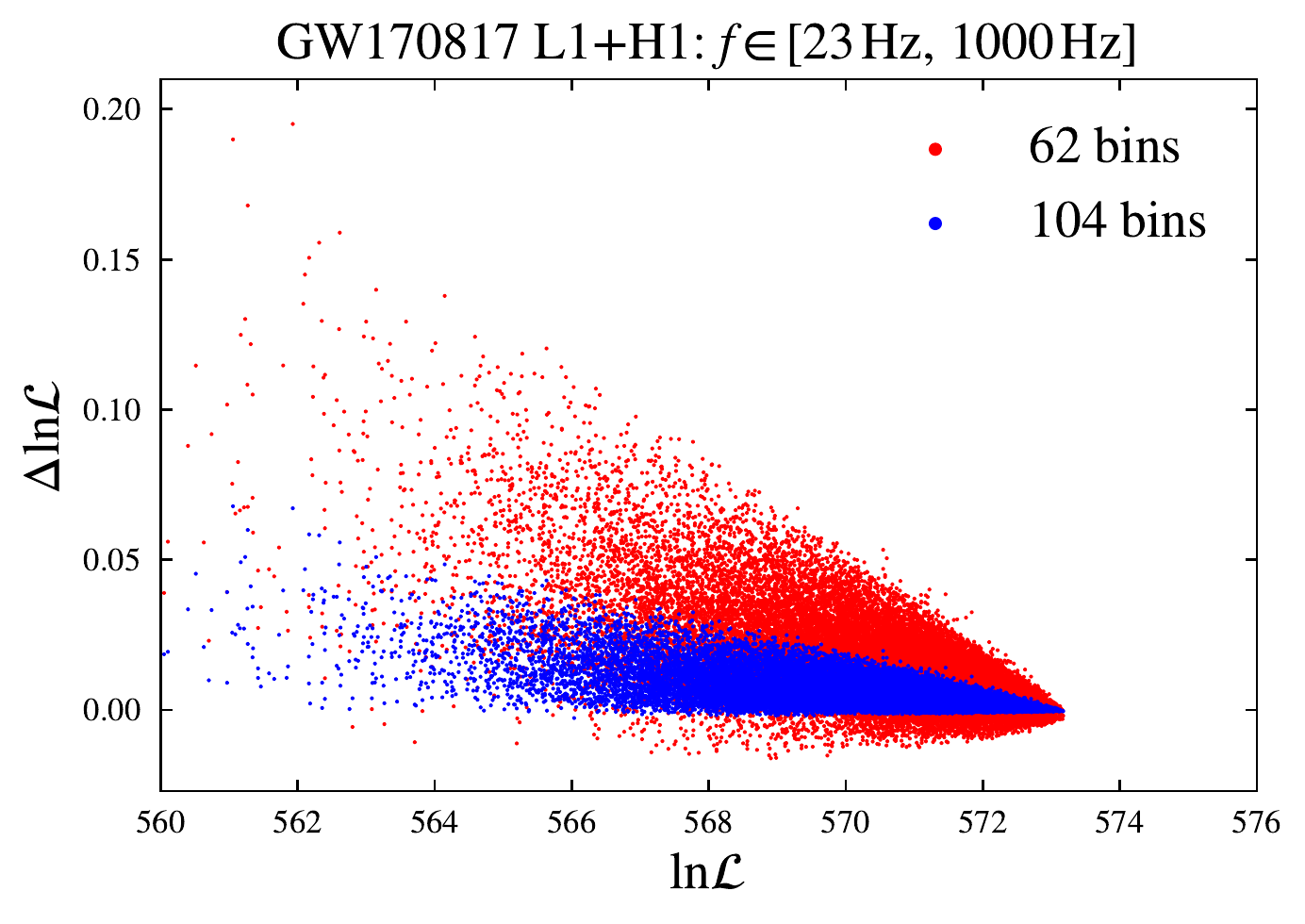}
  \caption{\label{fig:lnLerr} Absolute error in the evaluation of the log-likelihood $\Delta\ln\mathcal{L} = \ln\mathcal{L}_{\rm bin} - \ln\mathcal{L}_{\rm exact}$ due to relative binning. We compute the log-likelihood for the GW170817 data from both LIGO detectors, using frequencies from $23\,$Hz to $1000\,$Hz. The fiducial waveform $h_0$ used is a waveform very close but not equal to the best-fit solution. The dots are computed using posterior samples of parameter combinations from the output of the MCMC run. We show the quadratic improvement in the accuracy as the bin number roughly doubles from $62$ (red) to $104$ (blue).}
\end{center}
\end{figure}
%%%%%%%

We measure the accuracy of the relative binning method by computing the likelihood using both the binned summaries and the un-binned data.
Figure~\ref{fig:lnLerr} shows the error and the absolute value of the log-likelihood for $\approx 30000$ different parameter combinations that are sampled in the chains. We use common maximization routines to select a fiducial waveform $h_0$ that is very close, but not necessarily identical, to the best-fit one, $\ln\mathcal{L}_{\rm max} - \ln\mathcal{L}_{h_0} \approx 0.07$. For this reason, the error approaches zero near the maximum value of the log-likelihood $\ln\mathcal{L}_{\rm max} \approx 573.17$. Waveforms that deviate significantly from the best-fit one have lower absolute values of the log-likelihood and larger binning errors. For linear interpolation (as in \refeq{rf}), the relation happens to be linear because both the reduction in log-likelihood and the binning error scale quadratically with the magnitude of the perturbation. We can achieve the required accuracy of $\beta \lesssim 0.01$ in \refeq{like-err} using only 62 bins (using $\chi=1$ and $\epsilon=0.5\,{\rm rad}$). Figure~\ref{fig:lnLerr} also shows that we can increase the accuracy by using more bins (alternatively, we can use a higher-order interpolation scheme). 

{\it Simulating detector noise:} We can also use the binning method to efficiently simulate the effect of Gaussian detector noise without explicitly generating individual Fourier components. For this application, we replace the data stream $d(f)$ with $h_{\rm p}(f) + n(f)$, where $h_{\rm p}(f)$ is some injected physical waveform and $n(f)$ is the detector noise. The summary data $B_0(\bin)$ and $B_1(\bin)$ are not modified. We can pre-compute the contributions to the summary data $A_0(\bin)$ and $A_1(\bin)$ from the injection by replacing $d(f)$ with $h_{\rm p}(f)$ in \refeqs{A0}{A1}. Corrections to $A_0(\bin)$ and $A_1(\bin)$ due to $n(f)$ are themselves Gaussian random variables
\ba
 \delta A_0(\bin) & = & 4\,\sum_{f\in\bin}\,\frac{n(f)\,h^*_0(f)}{S_n(f)/T}, \\
 \delta A_1(\bin) & = & 4\,\sum_{f\in\bin}\,\frac{n(f)\,h^*_0(f)}{S_n(f)/T}\,\left( f - f_{\rm m}(b) \right).
 \ea
For given bin $\bin$, the two variables should be drawn from a bivariate normal distribution with zero mean and covariance:
 \ba
 && \VEV{\left|\delta A_0(\bin)\right|^2} = B_0(\bin), \quad \VEV{\delta A_0(\bin)\,\delta A^*_1(\bin)} = B_1(\bin), \nonumber\\
 && \VEV{\left|\delta A_1(\bin)\right|^2} = 4 \sum_{f\in\bin}\,\frac{|h_0(f)|^2}{S_n(f)/T}\,\left( f - f_{\rm m} \right)^2 \equiv B_2(\bin)\quad\quad.
\ea
In practice, this substantially simplifies feasibility studies for measurements using gravitational wave data.

%%%%%%%%%%%%%%%%%%%%%%%%%%%%%%%%%%%%%%%%%%%%%%%%%%%%%%%%%%%%%%%
%\section{Concluding Remarks}
%%%%%%%%%%%%%%%%%%%%%%%%%%%%%%%%%%%%%%%%%%%%%%%%%%%%%%%%%%%%%%%

{\it Concluding remarks: }We conclude that parameter estimation of gravitational wave mergers can be dramatically accelerated using relative binning. If we need to test a new waveform model against the data, we can readily obtain the likelihood from the precomputed summary data (provided we can easily compute the ratio between the test waveform and the fiducial waveform, see \refeq{rf}). A possible exception would be the case where the test waveform has sharp frequency-domain features that are unresolved by binning. 
Further, we anticipate that the importance of relative binning would increase with the duration of detected events, as expected with the advancement of the LIGO detectors and with the construction of LISA. 

% We make an additional point that further data compression is possible through a singular-value-decomposition of the relative waveforms $h(f)/h_0(f)$, which can be carried out either before or after the binning procedure. When applied to GW170817, we found that about $10$--$20$ components are sufficient to approximate all waveforms, but this comes at the price of some loss of generality, as the approximation basis depends on the waveform family.
Ultimately, in our method, we require two complex numbers per frequency bin for each waveform in order to compute the likelihood. We can further reduce the dimensionality of the binned relative waveforms using singular-value-decomposition. When applied to GW170817, we found that about $10$--$20$ components should be sufficient to approximate all waveforms. It may be possible to use this to further speed up the likelihood evaluation using reduced order quadrature (however, the approximation basis will then depend on the particular waveform family).

We comment that it is possible to implement relative binning also in time domain waveform models by working with the `complexified' waveforms $h_c(t) = h_{+}+ih_{\times}$ instead of the individual polarizations by computing $Z(d(f),h_c(f))$ using the inner product $\sum_t(\mathcal{F}^{-1}[d/S_n](t)h(t)$ (where $d/S_n$ is computed in the frequency domain, and $\mathcal{F}^{-1}$ is an inverse Fourier transformation).
We also comment that the binning technique can be generalized to account for non-quadrupolar modes, but this would require a different set of bins for each mode as different modes are highly oscillatory with respect to each other.

The relative binning method should facilitate fast gravitational wave data analysis without access to substantial computational resources. In addition, the algorithm we have presented is very elementary, and requires neither advanced knowledge of data sampling methods nor sophisticated basis-reduction pipelines. 

%for circular non-spinning binaries, and neglect the phase correction due to tidal deformation. We are not interested in the sky localization and source orientation, and hence treat the coalescence times $t_c$, the overall GW phases $\phi_c$, and the effective luminosity distances $D_{\rm eff}$ to be independent between the two LIGO detectors. we then maximize the likelihood with respect to these extrinsic parameters and explore the peak of the likelihood function in the two-dimensional parameter space $(\mathcal{M}^{\rm det}, \eta)$, where $\eta$ is the symmetric mass ratio. This suffices in order to demonstrate that the binning error is negligible in the interesting region of the likelihood surface.

%\appendix 

%%%%%%%%%%%%%%%%%%%%%%%%%%%%%%%%%%%%%%%%%%%%%%%%%%%%%%%%%%%%%%%%%%%%%%%%%%%%%%%%%%%

%\begin{acknowledgments}
\textit{Acknowledgments:} We thank Matias Zaldarriaga and Javier Roulet for helpful discussions during the writing of this paper, and Aaron Zimmerman for his insightful comments on an earlier draft. We also thank Hideyuki Tagoshi and Soichiro Morisaki for useful discussions.
BZ acknowledges support from the Infosys Membership Fund. LD is supported at the Institute for Advanced Study by NASA through Einstein Postdoctoral Fellowship grant number PF5-160135 awarded by the Chandra X-ray Center, which is operated by the Smithsonian Astrophysical Observatory for NASA under contract NAS8-03060. TV acknowledges support from the Schmidt Fellowship and the W.M. Keck Foundation Fund.
This research has made use of data, software and/or web tools obtained from the LIGO Open Science Center (https://losc.ligo.org), a service of LIGO Laboratory, the LIGO Scientific Collaboration and the Virgo Collaboration. LIGO is funded by the U.S. National Science Foundation. Virgo is funded by the French Centre National de Recherche Scientifique (CNRS), the Italian Istituto Nazionale della Fisica Nucleare (INFN) and the Dutch Nikhef, with contributions by Polish and Hungarian institutes.

%\end{acknowledgments}

%%%%%%%%%%%%%%%%%%%%%%%%%%%%%%%%%%%%%%%%%%%%%%%%%%%%%%%%%%%%%%%%%%%%%%%%%%%%%%%%%%

%------------------------------------------------------------------------------
\bibliographystyle{apsrev4-1-etal}
\bibliography{gw}

\end{document}